\documentclass[twocolumn]{aastex62}

\newcommand*\inst[1]{\unskip\hbox{\@textsuperscript{\normalfont$#1$}}}

\newcommand*\institute[1]{
  \begingroup
    \let\and\relax
    \renewcommand*\inst[1]{}%
    \renewcommand*\thanks[1]{}%
    \renewcommand*\email[1]{}%
  \endgroup
  \newcommand{\institutions}{#1}
}%

\let\oldarcsec\arcsec
\renewcommand\arcsec{\oldarcsec\xspace}%

\renewcommand{\ion}[2]{\textup{#1\,\textsc{\lowercase{#2}}}}

\usepackage{latexsym}		
\usepackage{graphicx}		
\usepackage{rotating}		
\usepackage{natbib}  
\usepackage{savesym}
\usepackage{amssymb}
\usepackage{amsmath}
\usepackage{morefloats}
\savesymbol{doublespace}
\usepackage{xspace}
\usepackage{color}
\usepackage{mdframed}
\usepackage{url}
\usepackage{subfigure}
\usepackage{grffile}
\usepackage{import}
\usepackage[utf8]{inputenc}
\usepackage{booktabs}
\usepackage{fancyhdr}

\usepackage[hang,flushmargin]{footmisc}
\usepackage{ifpdf}
\usepackage{microtype} 

\usepackage{times}

\newcommand{\msun}{\ensuremath{\mathrm{M}_{\odot}}\xspace}			

\newcommand{\hh}{\ensuremath{\textrm{H}_{2}}\xspace}			

\newcommand{\pc}{\ensuremath{\mathrm{pc}}\xspace}
\newcommand{\myr}{\ensuremath{\mathrm{Myr}}\xspace}

\newcommand{\hii}{\ion{H}{ii}\xspace}

\newcommand{\kms}{\textrm{km~s}\ensuremath{^{-1}}\xspace}	

\newcommand{\persc}{\ensuremath{\textrm{cm}^{-2}}\xspace}

\newcommand{\peryr}{\ensuremath{\textrm{yr}^{-1}}\xspace}

\def\ee#1{\ensuremath{\times10^{#1}}}

\def\eqref#1{Equation \ref{#1}}

\def\Figure#1#2#3#4#5{
\begin{figure*}[!htp]
\includegraphics[scale=#4,width=#5]{#1}
\caption{#2}
\label{#3}
\end{figure*}
}

\newenvironment{rotatepage}
{}{}

\begin{document}

\title{A HIGH CLUSTER FORMATION EFFICIENCY IN THE SAGITTARIUS~B2 COMPLEX}

\shorttitle{Cluster formation efficiency in Sgr~B2}
\shortauthors{Ginsburg \& Kruijssen}

\author[0000-0001-6431-9633]{Adam Ginsburg}
\affiliation{\it{Jansky fellow of the National Radio Astronomy Observatory, 1003 Lopezville Rd, Socorro, NM 87801 USA }}

\author[0000-0002-8804-0212]{J.~M.\ Diederik Kruijssen}
\affiliation{\textls[-10]{Astronomisches Rechen-Institut, Zentrum f\"{u}r Astronomie der Universit\"{a}t Heidelberg, M\"{o}nchhofstra{\ss}e 12-14, D-69120 Heidelberg, Germany}}

\correspondingauthor{Adam Ginsburg}
\email{aginsbur@nrao.edu}
\email{adam.g.ginsburg@gmail.com}

\begin{abstract}
    The fraction of stars forming in compact, gravitationally bound clusters
    (the `cluster formation efficiency' or CFE) is an important quantity for
    deriving the spatial clustering of stellar feedback and for tracing star
    formation using stellar clusters across the Universe.  Observations of
    clusters in nearby galaxies have revealed a strong dependence of the CFE on
    the local gas density, indicating that more stars form in star clusters
    when the star formation rate surface density is higher. Previously, it has
    not been possible to test this relation at very young ages and in clusters
    with individual stars resolved due to the universally-low densities in the
    cluster-forming regions in the Local Group. This has even led to the
    suggestion that the CFE increases with distance from the Sun, which would
    suggest an observational bias.  However, the Central Molecular Zone of the
    Milky Way hosts clouds with densities that are orders of magnitude higher
    than anywhere else in the Local Group.  We report a measurement of the CFE
    in the highest-density region in the Galaxy, Sgr~B2, based on ALMA
    observations of high-mass young stellar objects.  We find that over a
    third of the stars ($37\pm7\%$)  in Sgr~B2 are forming in bound clusters.
    This value is consistent with the predictions of environmentally-dependent
    models for the CFE and is inconsistent with a constant CFE in the
    Galaxy.
    \vspace{10mm}
\end{abstract}

\section{Introduction}
Gravitationally bound stellar clusters are some of the most important objects
in astronomy, providing both luminous probes of the star formation process at
great distances \citep[e.g.,][among many
others]{Brodie2006a,Adamo2013b,Kruijssen2018b,Kruijssen2018a} and large coeval
and co-located samples of stars in the local universe.  The prevalence of these
clusters varies substantially with environment: the fraction of star formation
occurring in bound, compact clusters, i.e.,~the \emph{cluster formation
efficiency} (CFE) $\Gamma$ is not constant
\citep{Adamo2015a,Johnson2016a,Messa2018a}.

\citet{Kruijssen2012a} proposed a theory in which $\Gamma$ is a function of gas
density\footnote{In galactic disks in hydrostatic equilibrium, this can be
expressed as the observable gas surface density, whereas in non-equilibrium
environments, the model depends on the gas volume density.}, with secondary
dependences on other global environmental quantities.  While this theory
reasonably explains observations spanning many galaxies, it has not yet been
directly tested in a
high-density environment where both the unclustered and clustered stars are
detected in a spatially resolved sense.  In this Letter, we perform such a test
in the Sgr~B2 cloud, a high-density region in the Galactic center in which both
stars and compact clusters (taken to be a proxy for gravitational boundedness)
are presently forming.

\section{Census of the Stellar Population}

We use the catalogs of millimeter and centimeter sources described in
\citet{Ginsburg2018a}, \citet{Gaume1995a}, and \citet{De-Pree2015a},
which consist of candidate young high-mass stars and young stellar objects
(YSOs), to infer the total stellar population in Sgr~B2.

\citet{Gaume1995a} observed Sgr~B2 at 1.3 cm with $\sim0.25$\arcsec resolution
with the VLA.  They detected 49 continuum sources.  These are exclusively \hii
regions and components of \hii regions.  \citet{De-Pree2015a} used 7 mm JVLA
Q-band observations at 0.05\arcsec resolution to catalog 26 sources in Sgr~B2~M
and 5 in Sgr~B2~N.  Of these, 7 detected in Sgr~B2~M were not reported in
\citet{Gaume1995a} because they were not resolved.  We assume each of the
VLA-detected sources is an \hii region and therefore contains at least one star
with $M\gtrsim20$~\msun, equivalent to a B0 star.

\citet{Ginsburg2018a} observed the cloud with ALMA, obtaining a resolution of
$0.5\arcsec$ in the 3 mm band.  They reported a total of 271 sources spread
throughout the cloud, of which 31 are confirmed \hii regions with implied
masses $M_*>20$~\msun; the rest are YSO candidates with
masses $M_*>8$~\msun.  We follow \citet{Ginsburg2018a} in extrapolating the total
stellar mass implied by the observed objects using a \citet{Kroupa2001a}
initial mass function.  With this extrapolation, each \hii region implies
the presence of a total of 326~\msun of stars, while each 8-20~\msun core
represents 135~\msun, assuming an upper stellar mass limit of 200~\msun (though
these numbers are weakly sensitive to the upper mass limit).

\section{Mass of the clusters}

In Table 2 of \citet{Ginsburg2018a}, four clusters are considered: N, M, NE, and
S.  Here, we re-evaluate the ``clusters" in NE and S.  These regions contain
few sources and are not centrally concentrated.
They are both moderate mass and, at present, are not guaranteed to form bound
clusters based on their irregular morphology.  We therefore exclude them from
the analysis, but note that if they are forming bound clusters, the measured
CFE would increase by a few percent. Our census of star and cluster formation
is summarized in Table~\ref{tab:clustermassestimates}.

\begin{table*}[htp]
\centering
\begin{minipage}{130mm}
\caption{Cluster Masses}
\begin{tabular}{cccccc}
\label{tab:clustermassestimates}
Name & $N({\rm cores})$ & $N({\rm H\textsc{ii}})$ & $M_{\rm inferred, cores}$ & $M_{\rm inferred, H\textsc{ii}}$ & $M_{\rm inferred,max}$ \\
 &  &  & $\mathrm{M_{\odot}}$ & $\mathrm{M_{\odot}}$ & $\mathrm{M_{\odot}}$ \\
\hline
M & 17 & 47 & 2300 & 15000 & 15000 \\
N & 11 & 3 & 1500 & 980 & 1500 \\
NE & 4 & 0 & 540 & 0 & 540 \\
S & 5 & 1 & 680 & 330 & 680 \\
Unassociated & 203 & 6 & 27000 & 2000 & 27000 \\
Total & 240 & 57 & 33000 & 19000 & 46000 \\
Clustered with NE, S & 37 & 51 & 5000 & 17000 & 18000 \\
Clustered only M, N & 28 & 50 & 3800 & 16000 & 17000 \\
\hline
\end{tabular}
\\
Partial reproduction of Table 2 in \citet{Ginsburg2018a}. $M_{\rm inferred,cores}$ and $M_{\rm inferred,\hii}$ are the inferred total stellar masses assuming the counted objects represent fractions of the total mass of 0.09 (cores) and 0.14 (\hii regions).  $M_{\rm inferred,max}$ is the greater of these two.  The \emph{Total} row represents the total over the whole observed region.  The two \emph{Clustered} rows show the total inferred mass of clusters including all four candidate clusters M, N, NE and S, then the mass of clusters including only M and N.
\end{minipage}
\end{table*}

\subsection{Cluster membership in Sgr~B2~M}
\label{sec:mmass}
\citet{Schmiedeke2016a} marked the Sgr~B2~M cluster as a 13\arcsec  (0.5 pc) radius
region centered on Sgr~B2~M~F3.  Within this volume, there are 47 \hii regions
in the joint  \hii region catalogs \citep{Gaume1995a,De-Pree2015a} from their
0.05\arcsec resolution 7 mm Q-band
VLA observations.  There are 17 non-\hii-region cores, the faintest of which is
1.3 mJy at 3 mm \citep{Ginsburg2018a}.  By extrapolating the \hii region counts,
\citet{Ginsburg2018a} inferred a total stellar mass of 1.5\ee{4}~\msun. The
implied volume density is $2.9\times10^4~\msun~\pc^{-3}$, which is a factor of
$\sim300$ above the local field star density
\citep{Launhardt2002a,Kruijssen2015a} and a factor of 50 higher than the mean
density of the circumnuclear gas stream in the Central Molecular Zone
\citep[CMZ; e.g.][]{Longmore2013b}. At such extreme densities, the free-fall
time is less than $0.1~\myr$, indicating that a compact morphology likely
indicates gravitational boundedness.

\citet{Schmiedeke2016a} give a gas mass of $M=9.6\ee{3}$~\msun in the Sgr~B2~M
cluster in their Table 2 and measure an instantaneous SFE of about 60\%, assuming
a cluster radius $r=0.5$ pc.  The total mass within 0.5 pc is then about $M_{\rm M} =
2.5\ee{4}$~\msun, and the escape speed is $v_{\rm esc}=14~\kms$. Again, this
high SFE (in combination with a low tidal filling factor of $r_{\rm h}/r_{\rm
t}<0.1$) indicates likely gravitational boundedness
\citep[e.g.][]{Baumgardt2007a,Kruijssen2012b}.

It is possible the Sgr~B2~M cluster is substantially larger, 35\arcsec (1.4 pc).
Within this radius, the `core' count is larger, 52 rather than 17, but the \hii
region count increases only marginally, from 47 to 49.  By contrast,
the gas mass is larger, $M_{\rm gas,1.4pc} = 8\ee{4}~\msun$, so the integrated SFE is lower,
about 20\%.  
The presence of many cores in the outskirts of the Sgr~B2~M cluster suggests
both that it may grow in stellar mass by accretion by up to an additional
$\sim50\%$ and that the lack of cores in the innermost region is due to
incompleteness (e.g., from confusion) rather than their absence, as suggested
in \citet{Ginsburg2018a}. However, the radial gradient of the SFE seems a
robust feature of Sgr~B2~M, which indicates that $r=0.5~\pc$ is the relevant
scale for a conservative identification of the bound cluster forming within the
system.

In conclusion, the stellar mass we adopt for the cluster, $M_{\rm star,M}=1.5\ee{4}$~\msun, is
a lower limit.  Adopting a higher value for $M_{\rm star,M}$, and correspondingly decreasing
the counted sources not associated with clusters, would increase the inferred
CFE by a few percent.

\subsection{Cluster membership in Sgr~B2~N}
\citet{Schmiedeke2016a} marked the Sgr~B2~N cluster as a 10\arcsec  (0.4 pc) radius circle
centered on Sgr~B2~N K2.  \citet{Schmiedeke2016a} identified 3 compact \hii regions
and \citet{Ginsburg2018a} identified 11 cores within this region.  The inferred
total stellar plus protostellar mass is 980-1500~\msun.  However, unlike Sgr~B2
M, Sgr~B2~N is gas-dominated, with $M_{\rm gas,N} = 2.8\ee{4}~\msun$ and
instantaneous SFE $\sim5\%$ \citep{Schmiedeke2016a}.  The escape speed from the
0.4 pc cluster is
$v_{\rm esc} = 18~\kms$.

The above mass and radius corresponds to a similar total volume density as Sgr
B2 M, even if the fraction in stars is considerably lower. Sgr~B2~N is
therefore better described as a `protocluster', in contrast with
Sgr~B2~M, which is a (very) young massive cluster (YMC).  Sgr~B2 N will need to
form an additional several thousand~\msun of stars to form a YMC, and will need
to do so at high efficiency.  However, since there is evidence that the
protocluster is still rapidly accreting both stars and gas, this outcome
is expected \citep[cf.][]{Ginsburg2016b}. Omitting the current stellar mass of
Sgr~B2~N from the total mass in bound clusters would decrease the inferred CFE
by 4\%.

\subsection{Velocity Dispersion Measurements - boundedness}
\label{sec:vdisp}
We measure the velocity dispersion of the stars as probed by their surrounding
\hii regions to confirm that the Sgr~B2~M cluster is presently gravitationally
bound; we do not have enough velocity measurements in Sgr~B2~N to perform a
similar measurement.
Because these regions are mostly hypercompact \hii regions with radii less than a few
thousand AU, they closely follow the motions of their stars and serve as 
probes of the underlying stellar kinematics.

We compare our velocity measurements to those of \citet{De-Pree2011a} and
\citet{De-Pree1996a} and perform  new velocity measurements based on the
data from \citet{Ginsburg2018a}.  Of the 32 unique \hii regions within the field
identified in the \citet{Gaume1995a} 1.3 cm data, 
15 had measurements in \citet{De-Pree2011a}.  We have measured an additional 11
velocities from the H41$\alpha$ radio recombination line.  Our measurements
agree to within 5 \kms with those of \citet{De-Pree2011a} for all sources we
both measured except F10.37, for which we measure a $\sim20~\kms$ discrepancy;
our spectrum is of much higher quality, so we adopt the H41$\alpha$ measurement
as correct.  All measured velocities are reported in Table \ref{tab:h41afits}.

We measure the 1D velocity dispersion in Sgr~B2~M by taking the standard
deviation of the measured V$_{\rm LSR}$ values.  Using only the
\citet{De-Pree2011a} measurements, we obtain $\sigma_{\rm 1D}\approx9~\kms$.  Using
the full data set, we obtain a higher $\sigma_{\rm 1D}\approx12~\kms$.  In both
cases, $\sigma_{\rm 1D}$ is  lower than the escape velocity
$v_{\rm esc}=14~\kms$ reported in Section \ref{sec:mmass}.

However, some individual sources are moving at high velocity with respect to
the average ($\bar{v}_{\rm LSR}(\rm H41\alpha) = 58.5~\kms$, $\bar{v}_{\rm LSR}(\rm H52\alpha)
= 65.8~\kms$), the fastest being G10.47 at $v_{\rm LSR}=34$~\kms or
$v_{\rm rel}=24-32$~\kms.  There is a small group at these highly negative
velocities and a projected distance from the center $r<0.1$ pc; these may be
bound to a deeper potential than we have inferred above, or they could be unbound
from the main cluster.
The \hii region J is separated by 0.4 pc and 16--24 \kms and is a diffuse \hii
region.  It may not be connected with the rest of the cluster.
If we exclude regions J, F10.37, G, G10.44, and G10.47, the velocity dispersion
drops to $\sigma_{\rm 1D}\approx8~\kms$.  If we exclude these sources, the total
inferred stellar mass for Sgr~B2~M drops from $M_{\rm star,M} = 1.5\ee{4}~\msun$ to
$M_{\rm star,M}=1.3\ee{4}~\msun$, corresponding to a drop of the CFE of 4\%.

The measurement of a velocity dispersion less than the escape speed in the cluster,
in conjunction with the high instantaneous SFE mentioned in Section \ref{sec:mmass},
implies that the Sgr~B2~M cluster is presently bound, and is strongly enough
bound that future gas loss is unlikely to unbind it.

\clearpage
\begin{table}[htp]
\caption{H41$\alpha$ Line Fits}
\begin{minipage}{130mm}
\begin{tabular}{llllllllllllllllll}
\label{tab:h41afits}
Source & Coordinates & $v_\mathrm{LSR}$(41) & $\sigma\left[v_\mathrm{LSR}(41)\right]$ & $\mathrm{FWHM}$(41) & $\sigma\left[\mathrm{FWHM}(41)\right]$ \\
 &  & $\mathrm{km\,s^{-1}}$ & $\mathrm{km\,s^{-1}}$ & $\mathrm{km\,s^{-1}}$ & $\mathrm{km\,s^{-1}}$ \\
\hline
A1 & 17:47:19.436 -28:23:01.36 & 62.7 & 0.5 & 31.6 & 1.1 \\
A2 & 17:47:19.566 -28:22:55.95 & 58.1 & 0.7 & 26 & 1.5 \\
B & 17:47:19.907 -28:23:02.91 & 75.6 & 0.4 & 34.9 & 0.9 \\
B10.06 & 17:47:19.868 -28:23:01.41 & 49.7 & 1.6 & 30.5 & 3.7 \\
B10.10 & 17:47:19.908 -28:23:02.13 & 70.1 & 1.7 & 27.3 & 3.9 \\
B9.96 & 17:47:19.776 -28:23:10.18 & 58.3 & 1.2 & 29.6 & 2.8 \\
B9.99 & 17:47:19.802 -28:23:06.9 & 61.7 & 0.8 & 23.2 & 1.8 \\
D & 17:47:20.053 -28:23:12.87 & 64.3 & 0.7 & 33.7 & 1.5 \\
E & 17:47:20.071 -28:23:08.65 & 61.3 & 0.3 & 29.3 & 0.8 \\
F1 & 17:47:20.12 -28:23:04.26 & 80.6 & 1.2 & 72.4 & 2.8 \\
F10.303 & 17:47:20.112 -28:23:03.7 & 57.2 & 1.1 & 81 & 2.7 \\
F10.33 & 17:47:20.14 -28:23:06.1 & 55.2 & 2.3 & 36.4 & 5.4 \\
F10.35 & 17:47:20.156 -28:23:06.73 & 58.7 & 7.1 & 72.2 & 17 \\
F10.37 & 17:47:20.179 -28:23:05.95 & 39.8 & 1 & 58 & 2.4 \\
F10.39 & 17:47:20.195 -28:23:06.65 & 63.1 & 1.4 & 53.5 & 3.3 \\
F2 & 17:47:20.17 -28:23:03.75 & 78.2 & 1.8 & 98.4 & 4.2 \\
F3 & 17:47:20.176 -28:23:04.81 & 61.1 & 0.4 & 45.2 & 1 \\
F4 & 17:47:20.219 -28:23:04.34 & 66.9 & 0.4 & 42.6 & 0.9 \\
G & 17:47:20.287 -28:23:03.07 & 44.3 & 0.5 & 43.8 & 1.3 \\
G10.44 & 17:47:20.246 -28:23:03.36 & 39.6 & 1.2 & 32.6 & 2.9 \\
G10.47 & 17:47:20.274 -28:23:02.38 & 34.1 & 1.6 & 22.4 & 3.7 \\
I & 17:47:20.511 -28:23:06.08 & 60.2 & 0.3 & 31.1 & 0.6 \\
I10.52 & 17:47:20.329 -28:23:08.14 & 60.9 & 2.1 & 22.8 & 4.9 \\
J & 17:47:20.574 -28:22:56.17 & 42.5 & 2 & 28.4 & 4.7 \\
\hline
\end{tabular}
\\
Velocities measured from radio recombination line fits for the HII regions in Sgr B2.  The H41$\alpha$ line comes from the ALMA data of \citet{Ginsburg2018a}.
\end{minipage}
\end{table}
\clearpage

\subsection{The cluster formation efficiency}
Table \ref{tab:clustermassestimates} shows the breakdown of ongoing star
formation within the Sgr~B2 region.  The total inferred mass of recently
formed or forming stars is ${M_{\rm *,total}\approx4.6\ee{4}~\msun}$ spread across
the whole cloud, with ${M_{\rm *,clustered}\approx1.7\ee{4}~\msun}$ concentrated
in the Sgr~B2~M and N clusters.  These values imply a CFE of
$\Gamma=M_{\rm *,clustered}/M_{\rm *,total}\times100\% = 37\%$. 

We have noted above that the membership of clusters M and N could be expanded,
and while this expansion would have no effect on the estimated mass of the clusters
(because their masses have been inferred from more complete samples of \hii regions),
it would reduce the number of unassociated cores by about 25\%, increasing the inferred
CFE to $\approx43\%$. By contrast, the omission of Sgr~B2~N and the
high-velocity \hii regions based on their possible unbound state would decrease
the CFE by up to 8\%. Given the similar magnitude of these two effects, we
adopt $\Gamma=37\pm7\%$. 

This measured CFE is substantially higher than measured in the Galactic disk.
\citet{Lada2003a} reported a CFE of $\Gamma=7^{+7}_{-3}\%$ for the local neighborhood, a
measurement $3$--$4\sigma$ below our lower value.  Our observation therefore demonstrates
that the CFE is non-uniform within the Galaxy.

\subsection{Comparison of observations to predictions}
\citet{Kruijssen2012a} described a theory for the CFE, in which the fraction
of stars forming in bound clusters ($f_{\rm bound}$) can be predicted based on both
`global' (galactic-scale, i.e., gas surface density, angular velocity, and
Toomre $Q$) and `local' (i.e., gas volume density, gas sound speed, gas
velocity dispersion) physical gas conditions. Using the observed parameters
listed in Table~\ref{tab:model}, we obtain predictions for $f_{\rm bound}$ and
compare them to the observed values of $\Gamma=37\pm7\%$. We obtain uncertainty
contours on the model predictions by carrying out a Monte Carlo error
propagation of the uncertainties on the observed parameters used as the input
values.

When evaluating the predicted CFE, we only consider the bound fraction of star formation
and omit the `cruel cradle effect' (CCE), i.e., the tidal disruption of
star-forming overdensities by dense substructure in the ISM. This choice is
made because the clouds on the CMZ dust ridge are moving coherently on a thin
stream, which limits the encounter rate relative to galactic discs.  The CCE
computed in \citet{Kruijssen2012a} assumed encounters with clouds could occur
in all three dimensions, so the timescales used in that model are not
appropriate for CMZ clouds.  Since we observe Sgr~B2 at a very early stage, the
clusters there have likely experienced few, if any, molecular cloud encounters
driving disruptive tidal shocks.

Figure \ref{fig:figure} shows the comparison of Sgr~B2 with other data and with
the theoretical prediction of the \citet{Kruijssen2012a} model. The model
predictions are slightly higher than the observed values, but they are
consistent within the expected errors. The comparison data sets are a
compilation from \citet{Adamo2015a}, plus spatially resolved M83 data
from the same paper (with gas surface densities from
\citealt{Freeman2017a}, as in \citealt{Reina-Campos2017a}).

The first panel in Figure \ref{fig:figure} shows the CFE $\Gamma$ as a function
of gas surface density, which is the fundamental dependence predicted by the
model. The data from this work on Sgr~B2 are shown as an orange point, and the
predictions from the \citet{Kruijssen2012a} model are overlaid in red and blue
contours for the global and local renditions of that theory, respectively. The
model and data are in clear agreement to within the uncertainties. The second
panel shows the CFE as a function of the star formation rate surface
density\footnote{Added after publication: The SFR surface density plotted is
the average over the dense gas `ring' that encompasses the dust ridge and Sgr
B2.  We adopt a CMZ SFR $\dot{M}=0.1$ \msun \peryr and an annulus of 60-120 pc
for the `ring', with $\pm20$ pc errorbars on the annulus radii, resulting in ${\Sigma_{SFR}=3_{-1}^{+6} \msun \mathrm{yr}^{-1} \mathrm{kpc}^{-2}}$.},
which is related to the gas surface density
\citep[e.g.,][]{Kennicutt1998a,Bigiel2008a,Leroy2013a} and is often considered
as the variable of interest in extragalactic studies of the CFE. The first two
panels show that Sgr~B2 fits along the relations defined by observations of
other galaxies. The third panel shows the CFE as a function of distance from
the Sun and is included to demonstrate that our measurement of the CFE in Sgr
B2 breaks the degeneracy between surface density and distance that existed in
previous studies. The distance to the object is not a predictor of the CFE (a
concern raised by \citealt{Adamo2011a} for extragalactic samples); instead, a
range of CFEs can be seen even within our own Galaxy.

\subsection{Effects of completeness}
The above analysis assumes the catalogs used are complete.  We note several
ways the survey may be incomplete, which were described in detail in
\citet{Ginsburg2018a}, and address the impact of any possible incompleteness
on our conclusions here.

The sample is complete above $M_*>20$~\msun. For the less massive `cores', the
completeness is much more difficult to evaluate, since models for the 3 mm
luminosity of such sources are limited.  For example, using the protostellar
evolution and radiative transfer models of \citet{Zhang2018b}, our survey is
100\% complete to any source down to 8~\msun and highly complete to
$\gtrsim5\msun$.  However, using the \citet{Robitaille2017a} models, our
completeness at 8-12~\msun ranges from 50-100\%, depending on the assumed
dust geometry around the central source.  However, if the catalog is less than 50\%
complete, the estimated star formation rate for Sgr~B2 would exceed
the total for the whole CMZ estimated through several independent methods
\citep{Barnes2017b}, suggesting that such low completeness is unlikely.

In the most extreme case, if the sources detected are all more massive than we
have assumed, the unclustered population would be larger relative to the clustered one.
We can establish a conservative lower limit on the CFE by directly comparing the
source counts, assuming that they all sample from the same population with a
common lower mass limit.  With 78 clustered out of a total of 297 sources, we
obtain a CFE of $\Gamma>26\%$.

The catalog has spatially variable completeness because of varying noise in the
images.  It is less complete in the clusters and their immediate vicinity.  The
effect of this incompleteness is to bias the cluster masses low relative to the
unclustered population.  However, the clusters are also well-populated by more
massive stars ($M_*>20~\msun$), to which the radio catalogs are very sensitive,
so this bias is unlikely to affect the results.

Finally, the definition of Sgr~B2 as a coherent region is somewhat arbitrary,
which propagates into our results in the form of a possible spatial
completeness.  The \citet{Ginsburg2018a} map covers a large area of about
$15\times15$ pc, whereas the region within the map that contains protostars is
$\sim5\times12$ pc. Surrounding this region, but still within the map, there is
a $>2$ pc `buffer' in which no protostars have been observed (see Figure 8 of
\citealt{Ginsburg2018a}), which strongly suggests that the sample used here
represents a single coherent star-forming region. If we apply the criterion
suggested by \citet{Alves2017b}, in which the `last' (lowest) closed contour in
a column density map defines the complete region, to the column density maps
used in \citet{Ginsburg2018a}, we find the observed region is complete to
column densities $N(\hh) > 2.5\ee{23}$~\persc, which is a factor of 2 below the
apparent threshold for star formation in both G0.253+0.016 and Sgr~B2 that was
noted by \citet{Ginsburg2018a}. This implies that our results are highly
unlikely to suffer from any spatial incompleteness.

\Figure{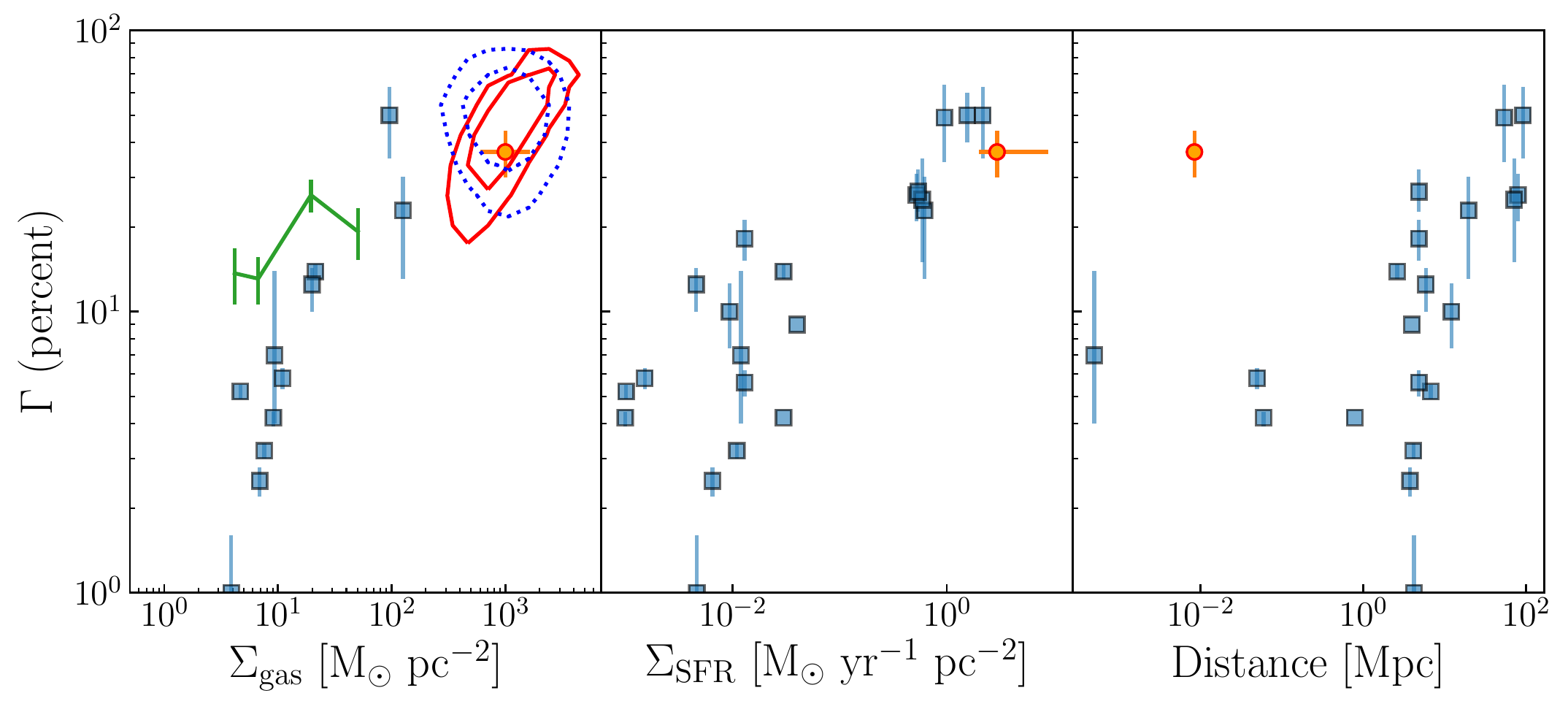}
{CFE as a function of gas surface density (left), star formation rate surface
density (middle), and distance from the Sun (right). The CFE of Sgr~B2 is
indicated with an orange symbol; literature observations of the CFE in
external galaxies are shown in blue. In the left-hand panel, the line indicates
the observed CFE as a function of galactocentric radius in M83
\citep{Adamo2015a}. The contours indicate the prediction of the
\citet{Kruijssen2012a} model for Sgr~B2 using the observed input parameters and
their uncertainties listed in Table~\ref{tab:model}. Contours are shown at the
95\% and 68\% confidence levels for the local (blue) and global (red)
renditions of the model. These panels demonstrate that there exist differences
in the CFE even within our Galaxy, and that these differences are consistent
with our theoretical understanding of bound cluster formation. 
}
{fig:figure}{1}{\textwidth}

\section{Conclusions}
We have measured the cluster formation efficiency in the Galactic Center cloud
Sgr~B2, resulting in $\Gamma=37\pm7\%$. This CFE is higher than that in the solar
neighborhood, implying that the CFE varies within the Milky Way. Specifically,
it changes with the galactic environment in a way that correlates with the
local gas conditions. This observation is consistent with existing extragalactic
observations.  However, it additionally rules out the idea that the
environmental dependence of the CFE is exclusively driven by an underlying
dependence on the distance from the Sun, which affected previous work and would have
been suggestive of an observational bias.  Instead, our results show that the
environmental variation of the CFE is a physical effect. The CFE model of
\citet{Kruijssen2012a}, in which higher average gas densities yield higher CFEs
due to shorter free-fall times and higher star formation efficiencies,
successfully predicts the observed value to within the uncertainties.

\begin{table*}[htp]
\centering
\begin{minipage}{130mm}
\caption{Model parameters}
\begin{tabular}{ccccccc}
\label{tab:model}
Quantity & Units & Median & Uncertainty & `Global model' & `Local model' & Reference \\
\hline
$\log{\Sigma}$ & [$\msun~\pc^{-2}$] & 3.00 & 0.22 & \checkmark &  & 4 \\
$\Omega$ & [$\myr^{-1}$] & 1.80 & 0.25 & \checkmark &  & 6,8 \\
$\log{\rho}$ & [$\msun~\pc^{-3}$] & 2.84 & 0.22 &  & \checkmark & 9 \\
$c_{\rm s}$ & [$\kms$] & 0.53 & 0.07 &  & \checkmark & 3,5 \\
$\log{\sigma}$ & [$\kms$] & 1.00 & 0.07 & \checkmark & \checkmark & 7 \\
$\log{\Sigma_{\rm GMC}}$ & [$\msun~\pc^{-2}$] & 3.61 & 0.18 & \checkmark & \checkmark & 2,10 \\
$\log{\alpha_{\rm vir}}$ & [--] & 0.04 & 0.16 & \checkmark & \checkmark & 10 \\
$\log{\beta_0}$ & [--] & $-0.47$ & 0.32 & \checkmark & \checkmark & 1,2 \\
$t_{\rm view}$ & [$\myr$] & 0.74 & 0.16 & \checkmark & \checkmark & 6 \\
\hline

$f_{\rm bound,global}$ & [\%] & 44.8 & 13.1 & \checkmark &  & this work \\
$f_{\rm bound,local}$ & [\%] & 48.9 & 11.6 &  & \checkmark & this work \\
\hline
\end{tabular}\\
\tablerefs{
(1) \citealt{Barnes2017b}, 
(2) \citealt{Federrath2016b}, 
(3) \citealt{Ginsburg2016a}, 
(4) \citealt{Henshaw2016b}, 
(5) \citealt{Krieger2017a}, 
(6) \citealt{Kruijssen2015a}, %
(7) \citealt{Kruijssen2018c}, %
(8) \citealt{Launhardt2002a}, 
(9) \citealt{Longmore2013b}, 
(10) \citealt{Walker2015a}. 
}
\end{minipage}
\end{table*}

\textit{Acknowledgements:}
We thank the anonymous referee for a timely and helpful report that led to
substantial improvement of the paper.
The National Radio Astronomy Observatory is a facility of the National Science
Foundation operated under cooperative agreement by Associated Universities,
Inc.
This paper makes use of the following ALMA data: ADS/JAO.ALMA\#2013.1.00269.S.
ALMA is a partnership of ESO (representing its member states), NSF (USA) and
NINS (Japan), together with NRC (Canada), NSC and ASIAA (Taiwan), and KASI
(Republic of Korea), in cooperation with the Republic of Chile. The Joint ALMA
Observatory is operated by ESO, AUI/NRAO and NAOJ.
JMDK gratefully acknowledges funding from the German
Research Foundation (DFG) in the form of an Emmy Noether Research Group (grant
number KR4801/1-1), from the European Research Council (ERC) under the European
Union's Horizon 2020 research and innovation programme via the ERC Starting
Grant MUSTANG (grant agreement number 714907), and from Sonderforschungsbereich
SFB 881 ``The Milky Way System'' (subproject P1) of the DFG.

\software{
The software used to make this version of the paper is available from github at
\url{https://github.com/keflavich/SgrB2_ALMA_3mm_Mosaic/} and
\url{https://github.com/keflavich/SgrB2_CFE}.  Pyspeckit \citep{Ginsburg2011c}
was used for the line fitting.  Plots were made with matplotlib \citep{Hunter2007a}.
}

\end{document}